\documentclass[12pt,a4paper]{article}
\usepackage[latin1]{inputenc}
\usepackage[all]{xy} 
\usepackage{graphicx} 
\usepackage[tight]{subfigure}
\usepackage{hyperref}
\usepackage[usenames,dvipsnames,svgnames,table]{xcolor}
\hypersetup{colorlinks=true, citecolor=Blue, linkcolor=Blue, urlcolor=Blue}
\usepackage{xcolor}
\usepackage{cancel}
\usepackage{lineno}

\usepackage{amsmath,amsfonts,amssymb,amsthm,epsfig,epstopdf,url,array}

\usepackage{todonotes}
\usepackage{graphicx}
\usepackage{color}

\def\be{\begin{equation}}
	\def\ee{\end{equation}}
\def\bi{\begin{itemize}}
	\def\ei{\end{itemize}}
\def\ed{\end{document}}

\newcommand{\R}{{\mathbb R}}

\renewcommand{\u}{{\bf u}}

\def\be{\begin{equation}}
	\def\ee{\end{equation}}

\theoremstyle{definition}
\newtheorem{defn}{Definition}[section]

\newtheorem{thm}{Theorem}[section]
\newtheorem{remark}{Remark}[section]

\newcommand{\q}{\quad}

\renewcommand{\u}{{\bf u}}

\begin{document}

\title{Nonlinear stability results for plane Couette and Poiseuille flows}

\author{Paolo Falsaperla, Andrea Giacobbe \& Giuseppe Mulone}
\date{}

\maketitle


\begin{abstract}
In this article we prove, choosing an appropriately weighted $L_2$-energy equivalent to the classical energy, that the plane Couette and Poiseuille flows are \textit{nonlinearly stable} with respect to \textit{streamwise perturbations}  for \textit{any Reynolds number}. In this case the \textit{coefficient of  time-decay} of the energy is $\pi^2/(2 {\rm Re})$, and it is a bound from above  of the time-decay of streamwise perturbations of \textit{linearized} equations. We also prove that the plane Couette and Poiseuille flows are nonlinearly stable if the Reynolds number is less then ${\rm Re}_{Orr}/\sin \varphi$ when the perturbation is a tilted perturbation, i.e.\ 2D perturbations with wave vector which forms an angle $\varphi \in [0, \pi/2]$ with the direction ${\bf i}$ of the motion. ${\rm Re}_{Orr}$ is the Orr (1907) critical Reynolds number for spanwise perturbations which, for the Couette flow is ${\rm Re}_{Orr}=177.22$ and for the Poiseuille flow is ${\rm Re}_{Orr}=175.31$.

In particular these results improve those obtained by Joseph (1966), who found for streamwise perturbations a critical nonlinear value of $82.6$ in the Couette case, and those obtained by Joseph and Carmi (1969), who found the value $99.1$ for plane Poiseuille flow for streamwise perturbations. The results we obtain here are, for any angle, in a good agreement with the experiments (see Prigent et al. 2003) and the numerical simulations (see Barckley and Tuckerman 2005, 2007).
\end{abstract}



\section{Introduction}

The study of stability and instability of the classical laminar flows of an incompressible fluid has attracted the attention of many authors for more that 150 years: Stokes, Taylor, Couette, Poiseuille, Kelvin, Reynolds, Lorentz, Orr, Sommerfeld, Squire, Joseph, Busse and many others.

This problem is nowadays object of study (see, for instance,   Deng and  Masmoudi 2018, Bedrossian et al. 2017,  Lan and Li 2013, Cherubini and De Palma 2013, Liefvendahl and Kreiss 2002) because the transition from laminar flows to instability, turbulence and chaos is not completely understood and there are some discrepancies between the linear and nonlinear analysis and the experiments (the so called \textit{Sommerfeld paradox 1908}).

The classical results are the following:

a) plane Poiseuille flow is \textit{linearly unstable} for ${\rm Re}>5772$ (Orszag 1971);

b) plane Couette flow and pipe Poiseuille flow are \textit{linearly stable} \textit{for all} Reynolds numbers (Romanov 1973);

c) in laboratory experiments plane and pipe Poiseuille flows undergo transition to three-dimensional turbulence for Reynolds numbers on the order of 1000. In the case of plane Couette flow the lowest Reynolds numbers at which turbulence can be produced and sustained has been shown to be between 300 and 400 both in the numerical simulations and in the experiments;

d) nonlinear asymptotic \textit{$L_2$-energy-stability} has been proved for Reynolds numbers ${\rm Re}$ below some value ${\rm Re}_E$ which is of the order $10^2$. In particular Joseph 1966 proved that ${\rm Re_E} ={\rm Re_E}^y=82.6$ (and ${\rm Re_E}^x=177.22$) for Couette flow, 
and Joseph and Carmi 1969 proved that ${\rm Re_E} ={\rm Re_E}^y=99.1$ (and ${\rm Re_E}^x=175.31$) for Poiseuille flow. 
Here and in what follows ${\rm  Re}^y$ refers to streamwise (or longitudinal) perturbations, 
${\rm Re}^x$ refers to spanwise (or transverse) perturbations.	

The use of \textit{weighted $L_2$-energy} has been fruitful for studying  nonlinear stability in fluid mechanics (see Straughan 2004).
Rionero and Mulone (1991) studied the non-linear stability of parallel shear flows with the Lyapunov method in the (ideal) case of stress-free boundary conditions. By using a weighted energy they proved that plane Couette flows and plane Poiseuille flows are conditionally asymptotically stable for all Reynolds numbers. 	

Kaiser et al. (2005) wrote the velocity field in terms of poloidal, toroidal and the mean field components. They used a generalized energy functional ${\cal E}$  (with some coupling parameters chosen in an optimal way) for \textit{plane Couette} flow, providing conditional nonlinear stability for Reynolds numbers ${\rm Re}$ below ${\rm Re_{\cal E}} = 177.22$, which is larger than the ordinary energy stability limit. The method allows the explicit calculation of so-called stability balls in the ${\cal E}$-norm; i.e., the system is stable with respect to any perturbation with ${\cal E}$-norm in this ball.

Kaiser and Mulone (2005) proved conditional nonlinear stability  for \textit{arbitrary plane parallel shear flows} up to some value ${\rm Re}_E$ which depends on the shear profile. They used a generalized (weighted) functional $ E $ and proved that ${\rm Re}_E$ turns out to be ${\rm Re}_E^x$, the ordinary energy stability limit for perturbations depending on $x$ (spanwise perturbations). In the case of the experimentally important profiles, viz. linear combinations of Couette and Poiseuille flow, this number is at least $175.31$, the value for pure Poiseuille flow. For Couette flow it is at least $177.22$.

Li and  Lin (2011) and Lan and Li (2013) gave a contribution towards the solution of the Sommerfeld paradox. They argue that even though the linear shear is linearly stable, slow orbits (also called quasi-steady states) in arbitrarily small neighbourhoods of the linear shear can be linearly unstable. They observe: ``The key is that in infinite dimensions, smallness in one norm does not mean smallness in all norms". Their study focuses upon a sequence of 2D oscillatory shears which are the Couette linear shear plus small amplitude and high spatial frequency sinusoidal shear perturbations.

Butler and  Farrell (1992) observed that transition to turbulence in plane channel flow occurs even for conditions under which modes of the linearized dynamical system associated with the flow are stable. By using variational methods they found linear three-dimensional perturbations that gain the most energy in a given time period.

Cherubini and De Palma (2013) used a variational procedure to identify nonlinear optimal disturbances in a Couette flow, defined as those initial perturbations yielding the largest energy growth at a short target time $T$, for given Reynolds number ${\rm Re}$ and initial energy $E_0$.

Recently, Bedrossian et al. (2017) studied Sobolev regularity disturbances to the periodic, plane Couette flow in 3D incompressible Navier-Stokes equations at high Reynolds number Re with the goal to estimate the stability threshold - the size of the largest ball around zero in a suitable Sobolev space $H^\sigma$ -  such that all solutions remain close to Couette. In particular, they proved the remarkable result: ``initial data that satisfies  $\Vert u_{in}\Vert_{H^\sigma} < \delta {\rm Re}^{-3/2}$ and some $\sigma>9/2$ depending only on $\sigma$  is global in time, remains
within $O({\rm Re}^{-1/2})$ of the Couette 
flow in $L^2$ for all time, and converges to
the class of ``2.5-dimensional" streamwise-independent solutions referred to
as \textit{streaks} for times $t \gtrsim {\rm Re}^{1/3}$".

Prigent et al. (2003) made experiments at the CEA-Sanclay Centre  to study, by decreasing the Reynolds number, the reverse transition from the turbulent to the laminar flow. At the beginning of their paper they wrote: ``In spite of more than a century of theoretical and experimental efforts, the transition to turbulence in some basic hydrodynamical flows is still far from being fully understood. This is especially true when linear and weakly nonlinear analysis cannot be used." The observed: ``a continuous transition towards a regular pattern made of periodically spaced, inclined stripes of well-defined width and alternating turbulence strength ... For lower ${\rm Re}$, a regular pattern is eventually reached after a transient during which domains, separated by wandering fronts, compete. The oblique stripes have a wavelength of the order of 50 times the gap. The pattern is stationary in the plane Couette flow case...The pattern was observed for $340 < {\rm Re} < 415$ in the plane Couette flow".

Barkley and Tuckerman (2007) studied numerically a turbulent-laminar ban\-ded pattern in plane Couette flow which is statistically steady and is oriented obliquely to the streamwise direction with a very large wavelength relative to the gap. They said: ``Regimes computed for a full range of angle and Reynolds number in a tilted rectangular periodic computational domain are presented ... The unusual but key feature of our study of turbulent-laminar patterns is the use of simulation domains aligned with the pattern wavevector and thus tilted relative to the streamwise-spanwise directions of the flow." For their numerical simulations they are guided by the experiments of Prigent et al. (2003). In their numerical simulation the domain is oriented such that ``the streamwise direction is tilted at angle $\theta= 24^\circ$ to the $x$-direction". In their Table 3 they, in particular,  reported turbulent-laminar ban\-ded patterns in plane Couette and plane Poiseuille flows. Parameters reported in Prigent et al. (2003) and in other papers are converted to a uniform Reynolds number based on the average shear and half-gap.  They show two columns that correspond to the values at the minimum and maximum Reynolds number reported. For plane Couette flow they found: ${\rm Re}=340$ for $\theta=37^\circ$, ${\rm Re}=395$ for $\theta=25^\circ$. For plane Poiseuille flow: ${\rm Re}=438$ for $\theta=19^\circ$, ${\rm Re}=357$ for $\theta=24^\circ$.

The aim of this paper is to show that a \textit{weighted $L_2$-energy} can be introduced to prove\textit{ nonlinear stability with respect to streamwise perturbations for any Reynolds number} and\textit{ nonlinear stability with respect to tilted perturbations}, which form an angle $\varphi$ with  the stream-direction ($x$-direction), for \textit{any Reynolds number less than the critical number $\dfrac{{\rm Re}_{Orr}}{\sin \varphi}$,} where ${\rm Re}_{Orr}$ is the Orr (1907) critical Reynolds number for spanwise perturbations (for the Couette and Poiseuille flows  ${\rm Re}_{Orr}=177.22$ and ${\rm Re}_{Orr}=175.31$, respectively). These results, in particular, confirm the results of Orr and improve those obtained by Joseph (1966) and Joseph and Carmi (1969).
Moreover, we note that our \textit{results are in a good agreement with the experiments} of Prigent et al. (2003) \textit{and the numerical computations} of Barkley and Tuckerman (2007).

The plan of the paper is the following. In Section 2 we recall the non-dimensional perturbation equations of laminar flows in a channel and we recall the classical linear stability/instability results. We study the linear stability with the classical $L_2$-energy and obtain the critical Reynolds numbers found by Orr (1907).
In Section 3 we first study the nonlinear stability of the laminar flows with respect to streamwise perturbations and prove that they are always stable, improving the results of Joseph (1966) and Joseph and Carmi (1969). In Subsection 3.2, we study the nonlinear stability with respect to tilted perturbations of an angle $\varphi$ with respect to the direction of the motion and prove that they are nonlinearly exponentially stable for any Reynolds number less that $\dfrac{{\rm Re}_{Orr}}{\sin\varphi}$ where  ${\rm Re}_{Orr}$ is the critical Reynolds number for spanwise perturbations.
The Section 3 ends with some remarks.

\section{Laminar flows between two parallel planes}

\begin{figure}{}{}
	\begin{center}
		\includegraphics[width=8cm]{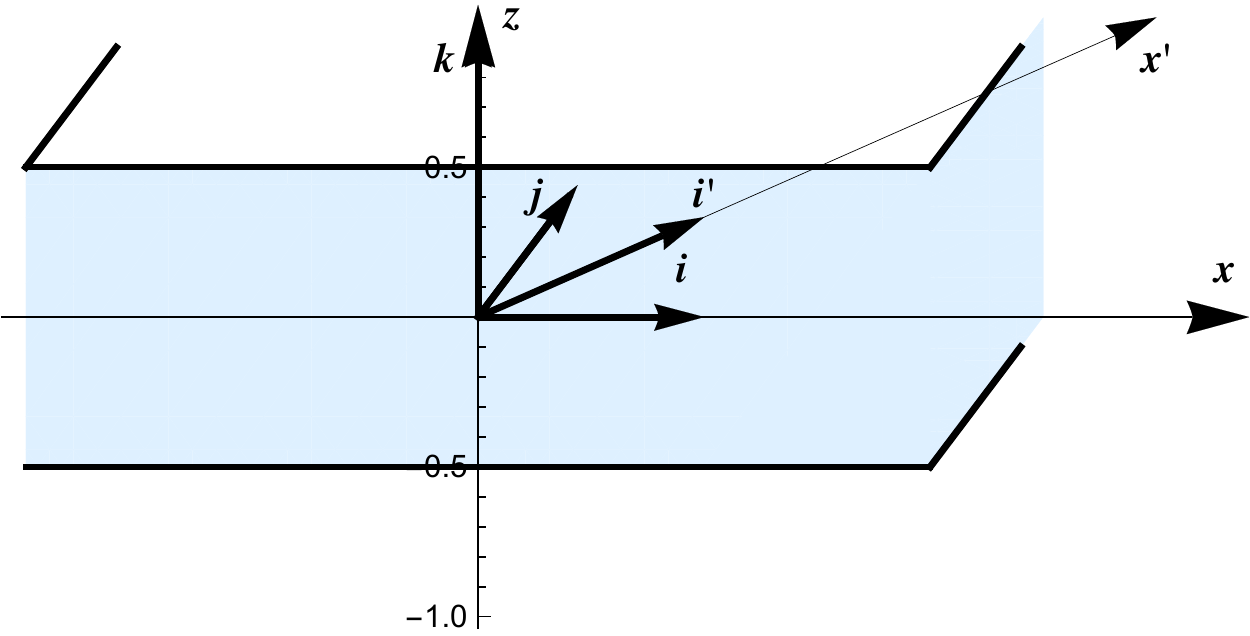}
	\end{center}  \caption{Laminar flows in a horizontal channel. The direction of motion is that of $x$-axis. The direction of $x'$-axis  forms an angle $\varphi$ with the direction of $x$-axis.}
\end{figure}

Consider, in a reference frame $Oxyz$, with unit vectors ${\bf i},{\bf j}, {\bf k}$, the layer $\mathcal D = \R^2 \times [-\frac{1}{2},\frac{1}{2}]$  of thickness $1$ with horizontal coordinates $x,y$ and vertical coordinate $z$.

Plane parallel shear flows, solutions of the Navier-Stokes equations, are characterized by the functional form

\be{\bf U}= \begin{pmatrix} \label{basic}

f(z)\\
0 \\
0
\end{pmatrix} =  f(z) {\bf i},
\ee

where ${\bf U}$ is the velocity field. 
The function $f(z) : [-\frac{1}{2} , \frac{1}{2}] \to \mathbb R$ is assumed to be sufficiently smooth and is called the shear profile. All the variables are written in a non-dimensional form.

In particular we have the well known profiles:

a) Couette  $f(z)=z$,

b) Poiseuille $f (z) = 1-4z^2$.

\subsection{Perturbation equations}

The perturbation equations to the plane parallel shear flows, in non-dimensional form, are 

\begin{eqnarray}\label{Couette-gen}
	\left\{ \begin{array}{l}
		u_t = -  {\bf u}\!\cdot\!\nabla u+ {\rm Re}^{-1} \Delta u -  (f u_x+f' w)- \dfrac{\partial p}{\partial x}\\[5pt]
		v_t = -  {\bf u}\!\cdot\!\nabla v+ {\rm Re}^{-1} \Delta v -  f v_x - \dfrac{\partial p}{\partial y}\\[5pt]
		w_t = -  {\bf u}\!\cdot\!\nabla w+  {\rm Re}^{-1}\Delta w -  f w_x- \dfrac{\partial p}{\partial z}\\	[5pt]	
		\nabla \cdot\u=0 ,\\
			\end{array}  \right.
\end{eqnarray}

where ${\bf u}= u{\bf i}+ v {\bf j}+w{\bf k}$ is the perturbation to the velocity field, $p$ is the perturbation to the pressure field and  ${\rm Re}$ is a Reynolds number.

To system (\ref{Couette-gen}) we append the rigid boundary conditions $${\bf u}(x,y,\pm 1/2,t)=0, \q (x,y,t) \in  \R^2 \times (0, +\infty) .$$

\begin{defn}
\textit{We define streamwise (or longitudinal) perturbations the perturbations ${\bf u}, p$  which do not depend on $x$. 
}\end{defn}

\begin{defn}
\textit{	We define spanwise (or transverse) perturbations the perturbations ${\bf u}, p$  which do not depend on $y$.} 
\end{defn}

We note that for spanwise perturbations  either $v \to 0$ exponentially fast as $t\to \infty$ or $v\equiv 0$.

\subsection{Linear stability/instability}

Assume that both ${\bf u}$ and $\nabla p $ are $x,y$-periodic with periods $\bar a$ and $\bar b$ in the $x$ and $y$ directions, respectively,  with wave numbers $(\bar a,\bar b) \in \R^2_+$  . In the following it suffices therefore to consider functions over the periodicity cell 
$$\Omega= [0, \frac{2\pi}{\bar a}]\times [0, \frac{2\pi}{\bar b}] \times [- \frac{1}{2}, \frac{1}{2}] .$$
As basic function space we take $L_2(\Omega)$. In the sequel $\Vert \cdot  \Vert$ represents  the norm in $L_2(\Omega)$. 

We recall that the classical results of Romanov (1973)  prove that Couette flow is \textit{linearly stable} for \textit{any Reynolds} number. Instead, Poiseuille flow is unstable for any Reynolds number bigger that $5772$ (Orszag, 1971).

We observe that, in the linear case, the Squire theorem (1933) holds and the most destabilizing perturbations are the spanwise bi-dimensional perturbations (see Drazin and Reid 2004). The critical  Reynolds value, for Poiseuille flow, can be obtained by solving the celebrated Orr-Sommerfeld equation (see Drazin and Reid 2004).

We also note that if we study the linear stability with the Lyapunov method, by using the classical energy 

$$V(t) = \dfrac{1}{2}[\Vert u \Vert^2 +  \Vert v \Vert^2 + \Vert w \Vert^2 ], $$
we obtain \textit{sufficient conditions of linear stability}.

By writing the energy identity

\be
\dot V= -(f'w,u) - {\rm Re}^{-1} [\Vert \nabla u \Vert^2+\Vert \nabla v \Vert^2+\Vert \nabla w \Vert^2], 
\ee
we have 
\be\label{stimV}
\begin{array}{l}
	\dot V= -(f'w,u) - {\rm Re}^{-1} [\Vert \nabla u \Vert^2+\Vert \nabla v \Vert^2+\Vert \nabla w \Vert^2] =\\[3mm]
	= \left(\dfrac{-(f'w,u)}{\Vert \nabla u \Vert^2+\Vert \nabla v \Vert^2+\Vert \nabla w \Vert^2} - \dfrac{1}{{\rm Re}}\right)\Vert \nabla {\bf u} \Vert^2  \le \\[3mm]
	\le \left( \dfrac{1}{\bar R} - \dfrac{1}{{\rm Re}}\right)\Vert \nabla {\bf u} \Vert^2 ,
\end{array}\ee 
where 
\be\label{maxRe}\dfrac{1}{\bar R}= \max_{\cal S} \dfrac{-(f'w,u)}{\Vert \nabla u \Vert^2+\Vert \nabla v\Vert^2+\Vert \nabla w \Vert^2}, \ee
and $\cal S$ is the space of the \textit{kinematically admissible fields } 
\be \begin{array}{l}\label{spaceS}
	{\cal S}= \{u, v, w \in H^1 (\Omega), \; u=v=w=0 \hbox{ on the boundaries},\\[3mm] \q u_x+v_y+w_z=0,\q  \Vert \nabla u \Vert+\Vert \nabla v \Vert+ \Vert \nabla w \Vert>0\},
\end{array}
\ee
and $ H^1(\Omega)$ is the Sobolev space of the functions which are in $L_2(\Omega)$ together with their first generalized derivatives.

The Euler-Lagrange equations of this maximum problem are given by 

\be\label{EL-Orr0}
\bar R (f' w {\bf i}+ f' u {\bf k}) - 2\Delta {\bf u} = \nabla \lambda,
\ee
where $\lambda$ is a Lagrange multiplier.

Since, for spanwise perturbations, $v\equiv 0$ and $\dfrac{\partial}{\partial y}\equiv 0$,   by taking the third  component of the \textit{double-curl} of  (\ref{EL-Orr0}) and by using the solenoidality condition $u_x+w_z=0$, 
we obtain the \textit{Orr equation} (1907)
\be
\dfrac{\bar R}{2} (f'' w_x+2f' w_{xz})+ \Delta \Delta w=0.	
\ee	

By solving this equation  we obtain the Orr results: for Couette and Poiseuille flows, we have ${\rm Re}_{Orr}= \bar R= 177.22$ (cf. Orr 1907 p. 128) and ${\rm Re}_{Orr}=\bar R= 175.31$ (cf. Drazin and Reid 2004 p. 163), respectively.

\section{Nonlinear stability}

Nonlinear stability conditions have been obtained by Orr (1907) in a celebrated paper, by using the Reynolds energy method (see Orr 1907 p. 122).

Orr writes: ``\textit{Analogy with other problems leads us to assume that disturbances in two dimensions will be less stable than those in three; this view is confirmed by the corresponding result in case viscosity is neglected}". He also says:  ``\textit{The three-dimensioned case was attempted, but it proved too difficult}".

Orr considers spanwise perturbations (i.e. $v\equiv 0$ and $\frac{\partial}{\partial y}\equiv 0$), the same perturbations that are used in the linear case by using Squire transformation (cf. Squire 1933, Drazin and Reid 2004).

The critical value he found, in the Couette Case,  is Re$^{x}=177.22$, where \textit{Re$^{x}$ is the critical Reynolds number with respect to spanwise perturbations} (see Orr 1907 p. 128, Joseph 1976 p. 181).

Joseph in his monograph (1976), pag. 181, says: ``\textit{Orr's assumption about the form of the disturbance which increases at the smallest {\rm Re} is not correct since we shall see that the energy of an $x$-independent disturbance (streamwise perturbations) can increase when  ${\rm Re} >2 \sqrt{1708}\simeq 82.65$}".

Joseph also gives a Table of values of the principal eigenvalues (critical energy Reynolds numbers) which depend on  a parameter  $\tau$ which varies from $0$ (streamwise perturbations) to $1$ (spanwise perturbations) and concludes that \textit{the value ${\rm Re} = 82.6$, 
is the limit for energy stability when $\tau=a=0$ (streamwise perturbations)}, where $a$ is the wave number in the $x$-direction.

However  \textit{we prove here that the conclusion of Joseph is not correct}. In fact, another ``weighted energy" (Lyapunov function) equivalent to the classical $L_2$-energy can be built that shows that the \textit{streamwise perturbations are always stable}. 
Moreover, we prove that the critical Ryenolds number for nonlinear stability with respect to ``tilted perturbations" (perturbations with axes parallel to a tilted $x'$-direction which form an angle $\varphi$ with the $x$-direction) are given by $\dfrac{{\rm Re}_{Orr}}{\sin\varphi}$.

This means that \textit{the results of Orr are correct} both for linear (as we have showed in Section 2) and  nonlinear stability for 2D perturbations (local nonlinear stability results up to Orr results,  are given in Kaiser et al. 2005, Kaiser and Mulone 2005 where the poloidal, toroidal and the mean flow representation of solenoidal vectors has been used and a suitable weighted energy has been introduced). 


\subsection{Streamwise perturbations}

First we assume that the perturbations are \textit{streamwise}, i.e. they \textit{do not depend on} $x$.

Therefore the perturbation equations become

\begin{eqnarray}\label{Couette-long}
	\left\{ \begin{array}{l}
		u_t = -  {\bf u}\!\cdot\!\nabla u+  {\rm Re}^{-1}\Delta u -  f' w\\	[5pt]	
		v_t = -  {\bf u}\!\cdot\!\nabla v+  {\rm Re}^{-1}\Delta v  - \dfrac{\partial p}{\partial y}\\	[5pt]	
		w_t = -  {\bf u}\!\cdot\!\nabla w+ {\rm Re}^{-1} \Delta w- \dfrac{\partial p}{\partial z}\\	[5pt]			
		v_y+w_z=0  .\\
			\end{array}  \right.
\end{eqnarray}

We define a \textit{weighted energy} (Lyapunov function) equivalent to the classical energy norm and show that the streamwise perturbations cannot destabilize the basic Couette or Poiseuille flows.

First we  introduce an  arbitrary positive number  $\beta$. Then, we multiply (\ref{Couette-long})$_1$ by $\beta u$ and integrate over $\Omega$. Besides, we multiply (\ref{Couette-long})$_2$ and  (\ref{Couette-long})$_3$ by $v$ and $w$ and integrate over $\Omega$. 
By taking into account of the solenoidality of ${\bf u}$, the boundary conditions and the periodicity, we have

$$\dfrac{d}{dt} [\dfrac{\beta \Vert u\Vert^2}{2}]= -\beta (f' w, u) - \beta {\rm Re}^{-1}\Vert \nabla u \Vert^2$$

$$\dfrac{d}{dt} [\dfrac{\Vert v\Vert^2+ \Vert w\Vert^2}{2}]=  - {\rm Re}^{-1} (\Vert \nabla v \Vert^2+ \Vert \nabla w )\Vert^2$$

By using the arithmetic-geometric mean inequality, we have

$$-\beta(f' w,u) \le \beta \dfrac{m^2}{2\epsilon} \Vert w \Vert^2 + \beta \dfrac{\epsilon}{2} \Vert u \Vert^2, $$

where $m= \max_{[-1/2,1/2]} \vert f'(z)\vert$ and $\epsilon$ is an arbitrary positive number.

Now we define  the Lyapunov function (weighted energy norm)

\be\label{energy-beta} E(t)= \dfrac{1}{2}[\beta \Vert u \Vert^2 +  \Vert v \Vert^2 + \Vert w \Vert^2 ], \ee

and choose $ \epsilon = \dfrac{\pi^2}{{\rm Re}}.$ From the above inequality and the use of the Poincar\'{e} inequality, 
we have

$$\dot E \le (\beta \dfrac{m^2 {\rm Re}}{2\pi^2}- \dfrac{\pi^2}{{\rm Re}})\Vert w\Vert^2 - \dfrac{\pi^2}{{\rm Re}} \Vert v \Vert^2 - \beta \dfrac{\pi^2}{2{\rm Re}} \Vert u \Vert^2,$$
and
$$\dot E \le (\beta \dfrac{m^2 {\rm Re}}{2\pi^2}- \dfrac{\pi^2}{{\rm Re}})\Vert w\Vert^2 - \dfrac{\pi^2}{2{\rm Re}} \Vert v \Vert^2 - \beta \dfrac{\pi^2}{2{\rm Re}} \Vert u \Vert^2.$$

By choosing $\beta = \dfrac{\pi^4}{m^2{\rm Re^2}},$ we finally have

$$ \dot E \le - \dfrac{\pi^2}{{\rm Re}} E.$$

Integrating this inequality, we have 
 the exponential decay

\be\label{expdecay} E(t) \le E(0) \exp\{-\dfrac{\pi^2}{{\rm Re}} t\}.\ee

This inequality implies global nonlinear exponential stability of the basic Couette or Poiseuille flows with respect to  the streamwise perturbations for any Reynolds number.
 
 Therefore we have proved: 
 
 \begin{thm}
 \textit{Assuming the perturbations to the basic shear flows (\ref{basic}) are streamwise, then we have nonlinear stability according to (\ref{expdecay}), with the energy $E$ given by (\ref{energy-beta}). 
} \end{thm}

From this Theorem we have 
 $ {\rm  Re}^y = +\infty$, i.e., \textit{the streamwise perturbations cannot destabilize the basic flows.}

\begin{remark}
	We note that this result is also in agreement with what observed Reddy et al. (1998) for the greatest  transient linear growth: ``\textit{for a given streamwise and spanwise wave number one can determine a disturbance, called an optimal, which yields the greatest transient linear growth. In channel flows, the optimals which yield the most disturbance growth are independent, or nearly independent, of the streamwise coordinate}".
\end{remark}

\begin{remark}
	We observe that the\textit{ time-decay coefficient we find, $\pi^2/(2 {\rm Re})$,  is in agreement} with the time-decay coefficient of longitudinal perturbations of \textit{linearized } equations,  see Figure 2.
	\end{remark}

\begin{figure}{}{}
	\begin{center}
		\includegraphics[width=8cm]{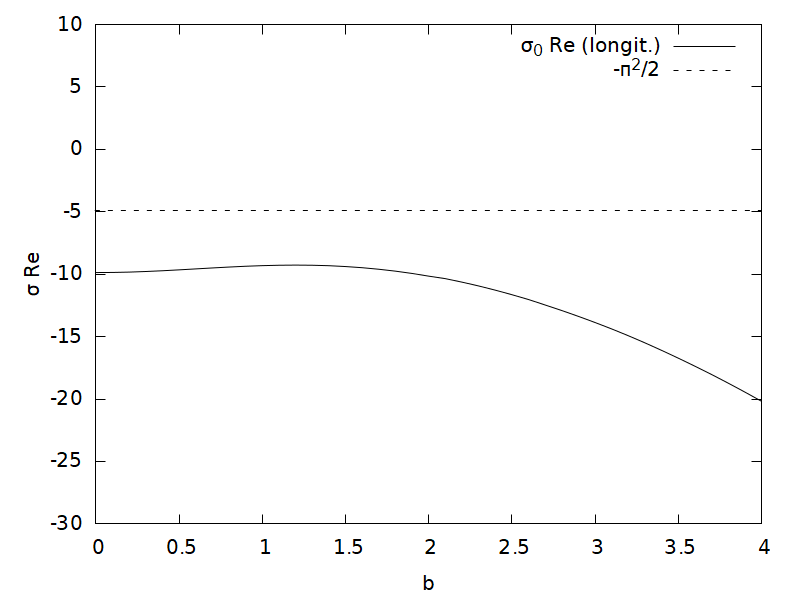}
	\end{center}  \caption{Dashed curve represents the square root of the energy time decay coefficient multiplied by ${\rm Re}$ (see (\ref{expdecay})). The continuous line represents the time decay coefficient of the most unstable streamwise linear perturbation $({\bf u}, p)$ as function of the wave number $b$}
\end{figure}

\subsection{Tilted perturbations}

\begin{defn}
	\textit{We define tilted perturbations of an angle $\varphi$ with  the $x$-direction the perturbations ${\bf u}, p$ along the $x'$ axis which do not depend on $x'$. 
}\end{defn}

Here we prove that, for 2D perturbations, the most destabilizing perturbations are the spanwise and the best stability results, in the energy norm, are those obtained by Orr (1907). Therefore, the results given by Joseph (1966) and by Joseph and Carmi (1969) are not the best because of a non optimal choice of an energy function.

For this, we consider an arbitrary tilted perturbation which forms an  angle $\varphi$ with the direction of motion ${\bf i}$ (the $x$-direction).

We have the following well-known relations


\be
\left\{\begin{array}{l}
	x'= \cos \varphi  \, x + \sin \varphi  \, y \\[2mm]
	y'= - \sin \varphi  \, x + \cos \varphi  \, y,
\end{array}\right. \q
\left\{\begin{array}{l}
	\dfrac{\partial }{\partial x'}= \cos \varphi  \, \dfrac{\partial }{\partial x} + \sin \varphi  \, \dfrac{\partial }{\partial y}\\[4mm]
		\dfrac{\partial }{\partial y'}= -\sin \varphi  \, \dfrac{\partial }{\partial x} + \cos \varphi  \, \dfrac{\partial }{\partial y}.
\end{array}\right. 
\ee

Moreover 

$${\bf u}= u {\bf i} + v {\bf j}+ w  {\bf k} = u' {\bf i'} + v' {\bf j'}+ w  {\bf k}, $$
with 

\be
\left\{\begin{array}{l}
	u'= \cos \varphi  \, u + \sin \varphi  \, v \\[2mm]
	v'= - \sin \varphi  \, u + \cos \varphi  \, v.
\end{array}\right. 
\ee

In the new variables, we have

$$\nabla' = (cos \varphi  \, \dfrac{\partial }{\partial x'} - \sin \varphi  \, \dfrac{\partial }{\partial y'}) {\bf i} + (\sin \varphi  \, \dfrac{\partial }{\partial x'} + \cos \varphi  \, \dfrac{\partial }{\partial y'}) {\bf j} +  \dfrac{\partial }{\partial z} {\bf k}, \q \Delta'=\Delta.$$

Consider the  perturbations system (\ref{Couette-gen}):

\begin{eqnarray}\label{Couette-gen1}
\left\{ \begin{array}{l}
u_t = -  {\bf u}\!\cdot\!\nabla u+ {\rm Re}^{-1} \Delta u -  (f u_x+f' w)- \dfrac{\partial p}{\partial x}\\[5pt]
v_t = -  {\bf u}\!\cdot\!\nabla v+ {\rm Re}^{-1} \Delta v -  f v_x - \dfrac{\partial p}{\partial y}\\[5pt]
w_t = -  {\bf u}\!\cdot\!\nabla w+  {\rm Re}^{-1}\Delta w -  f w_x- \dfrac{\partial p}{\partial z}\\	[5pt]	
\nabla \cdot\u=0 .\\
\end{array}  \right.
\end{eqnarray}

Multiply (\ref{Couette-gen1})$_1$ by $\cos \varphi$, (\ref{Couette-gen1})$_2$ by $\sin \varphi$, and add the equations so obtained. Besides, multiply (\ref{Couette-gen1})$_1$ by $-\sin \varphi$, (\ref{Couette-gen1})$_2$ by $\cos \varphi$, and add the equations so obtained.

Then we have the new system

 \begin{eqnarray}\label{Couette-gen2}
 \left\{ \begin{array}{l}
 u'_t = -  {\bf u}\!\cdot\!\nabla u'+ {\rm Re}^{-1} \Delta u' -  (f u'_x+f' \cos \varphi \,  w)- \dfrac{\partial p}{\partial x'}\\[5pt]
 v'_t = -  {\bf u}\!\cdot\!\nabla v'+ {\rm Re}^{-1} \Delta v' -  f v'_x + f' \sin \varphi \, w- \dfrac{\partial p}{\partial y'}\\[5pt]
 w_t = -  {\bf u}\!\cdot\!\nabla w+  {\rm Re}^{-1}\Delta w -  f w_x- \dfrac{\partial p}{\partial z}\\	[5pt]	
\dfrac{\partial u'}{\partial x'}+\dfrac{\partial v'}{\partial y'}+\dfrac{\partial w}{\partial z} =0 .\\
 \end{array}  \right.
 \end{eqnarray}

We note that when $\varphi \to 0$ then $x'\to x$, $y'\to y$, $u'\to u$, $v'\to v$.

Now we consider \textit{tilted perturbations in the $x'$-direction}, i.e, those with 	$\dfrac{\partial }{\partial x'}\equiv 0$. The first equation of  (\ref{Couette-gen2}) becomes

\be
 u'_t = -  {\bf u}\!\cdot\!\nabla u'+ {\rm Re}^{-1} \Delta u' -  (f u'_x+f' \cos \varphi \,  w) .
\ee

Proceedings as before,  we have the energy equation

\be \dfrac{d}{dt} [\dfrac{\beta \Vert u'\Vert^2}{2}]= -\beta (f' \cos\varphi u',w) - \beta {\rm Re}^{-1}\Vert \nabla u' \Vert^2, \ee
where $\beta$ is an arbitrary  positive number.
Moreover, we have

\be\label{eneqvw2} \dfrac{d}{dt} [\dfrac{\Vert v'\Vert^2+ \Vert w\Vert^2}{2}]=  - {\rm Re}^{-1} (\Vert \nabla v' \Vert^2+ \Vert \nabla w )\Vert^2+ (f' \sin \varphi \, v', w).\ee

We note that as $\varphi \to 0$ the energy equations tend to the energy equations obtained for the streamwise perturbations. Moreover, in the case $\dfrac{\partial }{\partial x'}\equiv 0$,  if $\varphi \to \dfrac{\pi}{2}$, since $y'\to -x$,  $v'\to -u$, $x'\to y$, 
$\dfrac{\partial v'}{\partial y'} \to \dfrac{\partial u}{\partial x}$, the energy equation (\ref{eneqvw2}) becomes that for spanwise perturbations.

Defining 

\be H =\dfrac{1}{2}[\Vert v' \Vert^2 + \Vert w \Vert^2 ], \ee
we have
\be\label{stimV}
\begin{array}{l}
	\dot H= (f' \sin \varphi \, v', w) - {\rm Re}^{-1} [\Vert \nabla v' \Vert^2+\Vert \nabla w \Vert^2] \le \\[3mm]
\le \left( \dfrac{1}{\bar R} - \dfrac{1}{{\rm Re}}\right)[\Vert \nabla v' \Vert^2+\Vert \nabla w \Vert^2],
\end{array}\ee 
where 
\be\label{maxRe}\dfrac{1}{\bar R}= \max_{\cal S} \dfrac{(f' \sin \varphi \, v',w)}{\Vert \nabla v'\Vert^2+\Vert \nabla w \Vert^2}, \ee
and $\cal S$ is the space of the \textit{kinematically admissible fields } 
\be \begin{array}{l}\label{spaceS}
	{\cal S}= \{v', w \in H^1 (\Omega), \; v'=w=0 \hbox{ on the boundaries},\\[3mm] \q v'_{y'}+w_z=0,\q  \Vert \nabla v' \Vert+ \Vert \nabla w \Vert>0\}.
\end{array}
\ee
The Euler-Lagrange equations for this maximum problem are
\be\bar R \sin \varphi (f' w {\bf j'}+ f' v' {\bf k}) +2 (\Delta v' {\bf j'}+ \Delta w {\bf k})= \nabla \mu, \ee
where $\mu$ is a Lagrange multiplier.
By taking the third component of the \textit{double-curl} of this equation, we have 

\be \dfrac{\bar R \sin \varphi}{2} (f'' w_{y'}+ f' w_{y'z}- f' v'_{y'y'})-\Delta \Delta w=0.\ee
Since, by solenoidality condition $\dfrac{\partial v'}{\partial y'} = - \dfrac{\partial w}{\partial z}$, we obtain

\be \dfrac{\bar R \sin \varphi}{2} (f'' w_{y'}+ 2f' w_{y'z})-\Delta \Delta w=0.\ee

By defining $\tilde y= -y'$ the operator $\Delta$ does not change and we have 
\be \dfrac{\bar R \sin \varphi}{2} (f'' w_{\tilde y}+ 2f' w_{\tilde y z})+ \Delta\Delta w=0.\ee

This equation coincides with the celebrated Orr equation if we substitute the critical Reynolds number  ${\rm Re}_{Orr}$ in that equation with $\bar R \sin \varphi$. Thus we obtain as critical nonlinear Reynolds number for $x'$-independent perturbations the critical value obtained by Orr divided by $\sin \varphi$.

In particular we note that for $\varphi \to \frac{\pi}{2}$ we obtain the critical Orr Reynolds number for spanwise perturbations. For $\varphi \to 0$ we obtain, as we have showed before, that the critical Reynolds number tends to $+ \infty$.

Now we have to prove that also the component $u'$ tends to zero as $t \to \infty$. For this, supposing that ${\rm Re}< \bar R$ and  by using the Poincaré inequality, we have 

\be 
\dot{H} \le  \left( \dfrac{1}{\bar R} - \dfrac{1}{{\rm Re}}\right)  \pi^2 \,  [\Vert v' \Vert^2+ \Vert w \Vert^2] = -2r\pi^2 H,
\ee
where $$r=  \left( \dfrac{1}{{\rm Re}}- \dfrac{1}{\bar R} \right).$$

 By defining 

\be\label{energob} E(t)= \dfrac{1}{2}[\Vert v' \Vert^2 + \Vert w \Vert^2 ]+ \dfrac{\beta \Vert u' \Vert^2}{2} , \q \beta >0. \ee

The energy equation is

\be
\begin{array}{l}
\dot E = -\beta {\rm Re}^{-1} \Vert \nabla u' \Vert^2 - \beta (f' \cos \varphi \, u', w) - 
	{\rm Re}^{-1} (\Vert \nabla v' \Vert^2 + \Vert \nabla w \Vert^2)+\\ \q + (f' \sin \varphi \, v', w).
\end{array}
\ee 
We have the following estimate

\be\label{stim11}
\begin{array}{l}
\dot E \le	-r\pi^2 (\Vert v' \Vert^2 + \Vert  w \Vert^2) + m \beta \Vert w \Vert \Vert u' \Vert - \beta {\rm Re}^{-1} \pi^2 \Vert u' \Vert^2,
\end{array}\ee 
where \be\label{maxm} m= \max_{[-1/2,1/2]} \vert f'(z)\vert \cos \varphi.\ee
We first note that if $\varphi =\frac{\pi}{2}$, then the right hand side of the previous inequality is always negative for $r>0$. We consider the case of $\varphi <\frac{\pi}{2}$.

By arithmetic-geometric mean inequality, we have 

\be m \beta \Vert u' \Vert \Vert w \Vert \le \dfrac{\beta m^2}{2 \epsilon} \Vert w \Vert^2 +  \dfrac{\beta \epsilon}{2}  \Vert u' \Vert^2. \ee

Therefore
\be\label{stim12}
\begin{array}{l}
	\dot E \le	 ( \dfrac{\beta m^2}{2 \epsilon} -r\pi^2)\Vert  w \Vert^2 -  r \pi^2  \Vert v' \Vert^2 + \beta (\dfrac{\epsilon}{2}- {\rm Re}^{-1} \pi^2) \Vert u' \Vert^2.
\end{array}\ee 
By choosing $\epsilon= \dfrac{\pi^2}{\rm Re}$ and $\beta= \dfrac{r\pi^4}{m^2 {\rm Re}}$, we obtain
\be 
\dot E \le -\dfrac{r\pi^2}{2} (\Vert w \Vert^2+\Vert v' \Vert^2 + \beta \Vert u' \Vert^2) = -r \pi^2 E
\ee
Assuming ${\rm Re} < \bar R$, i.e., $r>0$, we finally have

\be\label{enob}E(t) \le E(0) e^{-r \pi^2 t}.\ee

Therefore we have proved 

\begin{thm}
	\textit{Assuming the perturbations to the basic shear flows (\ref{basic}) are tilted of an angle $\varphi$, then we have nonlinear stability according to (\ref{enob}), with the energy $E$ given by (\ref{energob}). 
} \end{thm}

For the sake of completeness, in Table 1 we report the critical Reynolds numbers obtained, for Couette and Poiseuille flows, for some inclination angle $\varphi$ with respect to $x$-direction.

{\scriptsize 

\begin{table} \label{table}
	\begin{center}
		\begin{tabular}{|l|l|l|l|l|l|l|}
			\hline
			
			Inclination	& Couette & Poiseuille  \\[7pt]
\hline
$\varphi=0^\circ$ & $+\infty$  & $+\infty$  \\[7pt]
\hline
$\varphi=1^\circ$ & $10154.5$  & $10045$  \\[7pt]
\hline
$\varphi=3^\circ$ & $3386.2$  & $3349.7$  \\[7pt]
\hline
$\varphi=5^\circ$ & $2033.37$  & $2011.46$  \\[7pt]
\hline
$\varphi=10^\circ$ & $1020.57 $  & $1009.57$  \\[7pt]
\hline
$\varphi=15^\circ$ & $684.72 $  & $677.35$  \\[7pt]
\hline
$\varphi=20^\circ$ & $ 518.16$  & $512.57$  \\[7pt]
\hline
$\varphi=24^\circ$ & $ 435.71$  & $431.02$  \\[7pt]
\hline
$\varphi=30^\circ$ & $ 354.44$  & $350.62$  \\[7pt]
\hline
$\varphi=40^\circ$ & $ 275.71$  & $272.73$  \\[7pt]
\hline
$\varphi=50^\circ$ & $ 231.34$  & $228.85$  \\[7pt]
\hline
$\varphi=60^\circ$ & $204.64 $  & $202.43$  \\[7pt]
\hline
$\varphi=70^\circ$ & $ 188.59$  & $186.56$  \\[7pt]
\hline
$\varphi=80^\circ$ & $ 179.95$  & $178.01$  \\[7pt]
\hline
$\varphi=90^\circ$ & $ 177.22$  & $ 175.31$  \\[7pt]
\hline								
\end{tabular}		
\end{center} \caption{Critical Reynolds numbers for 2D perturbations with respect to tilted $x'$-perturbations which do not depend on $x'$, for different inclination angles $\varphi$. The streamwise perturbations correspond to $\varphi=0^\circ$, the spanwise perturbations correspond to $\varphi=90^\circ$.}
\end{table}
}

\begin{remark}
	We note that the results we have obtained here continue to hold for arbitrary plane parallel shear flows. For this it is sufficient to compute the maximum $m$ given by (\ref{maxm}).
\end{remark}

\begin{remark}
	From Table 1, we see that, in Couette case, the critical values of tilted perturbations nearest the experimental data are those in a neighbourhood of $30^\circ$ (roughly the interval  $\varphi \in [24^\circ, 40^\circ]$). In the case of Poiseuille flows, the critical values nearest to the experiments are those  in a neighbourhood of $\varphi=10^\circ$.
	In particular, in the numerical simulation of Barkley and Tuckerman (2007) the domain is oriented such that ``the streamwise direction is tilted at angle $\theta= 24^\circ$ to the $x$-direction". As we have observed in the Introduction, they found that the pattern is stationary in
	the plane Couette flow case...The pattern was observed for $340 < {\rm Re} < 415$ in the plane Couette flow". The critical Reynolds value given by Joseph (1976) for $\varphi=24^\circ$ (in this case the value $\tau$ of Joseph is $\tau=0.40$) is ${\rm Re}=88.57$.  Instead our result gives the critical value ${\rm Re}=435.71$. 
	
\end{remark}

\begin{remark}
The method used here can be applied also to magnetohydrodynamics Couette and Hartmann
shear flows. This will be done in a future work.		
\end{remark}

\begin{remark}
We believe that the results obtained in the present paper make a  contribution to the solution of the Sommerfeld paradox.	
\end{remark}

\vskip .3cm

\textit{Acknowledgements}.  The research that led to the present paper was partially supported by a grant of the group GNFM of INdAM and by a grant: PTR-DMI-53722122113``Analisi qualitativa per sistemi dinamici finito e infinito dimensionali
con applicazioni a biomatematica, meccanica dei continui e termodinamica estesa classica e quantistica" of University of Catania. The authors acknowledge also support from the project PON SCN 00451 CLARA - CLoud plAtform and smart underground imaging for natural Risk Assessment, Smart Cities and Communities and Social Innovation.

\phantomsection
\vspace*{0.5cm}
{\footnotesize\begin{tabular}{rl}
\hline
& \\
\label{A}$^{a}$
& Universit{\`{a}} degli Studi di Catania\\
& Dipartimento di Matematica e Informatica\\
& Viale A. Doria 6\\
& 9512500 Catania, Italy\\
& E-mail: 
\href{mailto:falsaperla@dmi.unict.it}{falsaperla@dmi.unict.it} \href{mailto:giacobbe@dmi.unict.it}{giacobbe@dmi.unict.it}, \href{mailto:giuseppe.mulone@unict.it}{giuseppe.mulone@unict.it}\\
\end{tabular}}

\end{document}